\documentclass[]{aastex701}

\usepackage{amsmath}

\begin{document}

\title{SN 2007it on the RISE - a radio detection of an interacting supernova 18 years post-explosion}

\author[0000-0002-6606-2816]{F. Acero}
\affiliation{FSLAC IRL 2009, CNRS/IAC, La Laguna, Spain}
\email[]{} 

\author[0000-0001-5609-7372]{R. Z. E. Alsaberi}
\affiliation{Western Sydney University, Locked Bag 1797, Penrith South DC, NSW 2751, Australia}
\affiliation{Faculty of Engineering, Gifu University, 1-1 Yanagido, Gifu 501-1193, Japan}
\email[]{} 

\author[0000-0002-7918-904X]{M. Arias}
\affiliation{Instutito de Astrof\'isica de Andaluc\'ia IAA-CSIC, Glorieta de la Astronom\'ia s/n, 18008 Granada, Spain}
\email[]{} 

\author[]{J. Borowska-Naguszewska}
\affiliation{Humboldt-Universität zu Berlin, Faculty of Mathematics and Natural Sciences, Newtonstraße 15, 12489 Berlin, Germany}
\email[]{} 

\author[0000-0002-8312-6930]{R. Brose}\thanks{Corresponding author: \href{mailto:robert.brose@mail.de}{robert.brose@desy.de}}
\affiliation{Institute of Physics and Astronomy, University of Potsdam, 14476 Potsdam-Golm, Germany}
\email[]{robert.brose@desy.de} 

\author[0000-0002-7239-2248]{C. Burger-Scheidlin}
\affiliation{Astronomy \& Astrophysics Section, School of Cosmic Physics, Dublin Institute for Advanced Studies, DIAS Dunsink Observatory, Dublin D15 XR2R, Ireland} 
\affiliation{School of Physics, University College Dublin, Belfield, Dublin 4, Ireland}
\email[]{}

\author[0000-0002-8186-4753]{P. G. Edwards}
\affiliation{CSIRO Space and Astronomy, Australia Telescope National Facility, PO Box 76, Epping, NSW 1710, Australia}
\email[]{} 

\author[0000-0001-6674-4238]{Q. Feng}
\affiliation{Department of Physics \& Astronomy, University of Utah, Salt Lake City, UT 84112, USA}
\email[]{} 

\author[0000-0002-4990-9288]{M. D. Filipovi\'c}
\affiliation{Western Sydney University, Locked Bag 1797, Penrith South DC, NSW 2751, Australia}
\email[]{} 

\author[0000-0003-1792-2338]{T. Laskar}
\affiliation{Department of Physics \& Astronomy, University of Utah, Salt Lake City, UT 84112, USA}
\email[]{} 

\author[0000-0001-6109-8548]{S. Lazarevi\'c}\thanks{Corresponding author: \href{mailto:s.lazarevic@westernsydney.edu.au}{s.lazarevic@westernsydney.edu.au}}
\affiliation{Western Sydney University, Locked Bag 1797, Penrith South DC, NSW 2751, Australia}
\affiliation{CSIRO Space and Astronomy, Australia Telescope National Facility, PO Box 76, Epping, NSW 1710, Australia}
\affiliation{Astronomical Observatory, Volgina 7, 11060 Belgrade, Serbia}
\email[]{s.lazarevic@westernsydney.edu.au} 

\author[0000-0002-5449-6131]{J.~Mackey}
\affiliation{Astronomy \& Astrophysics Section, School of Cosmic Physics, Dublin Institute for Advanced Studies, DIAS Dunsink Observatory, Dublin D15 XR2R, Ireland} 
\affiliation{School of Physics, University College Dublin, Belfield, Dublin 4, Ireland}
\email[]{} 

\author[0009-0005-9192-5491]{A. Nucara}
\affiliation{INAF - IAPS, Via Fosso del Cavaliere 100, I-00033 Rome, Italy}
\affiliation{Università Tor Vergata, Dipartimento di Fisica, Via della Ricerca Scientifica 1, I-00133 Rome, Italy}
\email[]{} 

\author[0000-0002-7329-3209]{K. Rose}\thanks{Corresponding author: \href{mailto:kovi.rose@sydney.edu.au}{kovi.rose@sydney.edu.au}}
\affiliation{Sydney Institute for Astronomy, School of Physics, The University of Sydney, Camperdown, NSW 2006, Australia} 
\affiliation{CSIRO Space and Astronomy, Australia Telescope National Facility, PO Box 76, Epping, NSW 1710, Australia}
\email[]{kovi.rose@sydney.edu.au} 

\author[0000-0003-4501-8100]{S. Ryder}
\affiliation{School of Mathematical and Physical Sciences, Macquarie University, Sydney, NSW, Australia} 
\affiliation{Astrophysics and Space Technologies Research Centre, Macquarie University, Sydney, NSW, Australia}
\email[]{} 

\author[0000-0003-1500-6571]{F. Schüssler}
\affiliation{Institute for Research on the Fundamental Laws of the Universe (IRFU), Commissariat à l’énergie atomique (CEA), Université Paris-Saclay, F-91191 Gif-sur-Yvette, France.}
\email[]{} 

\author[0009-0000-3416-9865]{A. Simongini}
\affiliation{INAF - Osservatorio Astronomico di Roma, Via di Frascati 33, I-00078 Monteporzio Catone, Italy}
\affiliation{Università Tor Vergata, Dipartimento di Fisica, Via della Ricerca Scientifica 1, I-00133 Rome, Italy}
\email[]{} 

\author[0009-0009-7061-0553]{Z. J. Smeaton}
\affiliation{Western Sydney University, Locked Bag 1797, Penrith South DC, NSW 2751, Australia}
\email[]{} 

\author[0000-0002-2814-1257]{I. Sushch}
\affiliation{Centro de Investigaciones Energ\'eticas, Medioambientales y Tecnol\'ogicas (CIEMAT), E-28040 Madrid, Spain}
\affiliation{Astronomical Observatory of Ivan Franko National University of Lviv, Kyryla i Methodia 8, 79005 Lviv, Ukraine}
\affiliation{Centre for Space Research, North-West University, 2520 Potchefstroom, South Africa}
\email[]{} 

\author[0000-0002-6468-8292]{S. Zhu}
\affiliation{DESY, D-15738 Zeuthen, Germany}
\email[]{} 

\collaboration{all}{The RISE collaboration}

\begin{abstract}

We report the first detection of radio emission from the Type\,II supernova SN\,2007it, located at a distance of $12.2\,$Mpc in NGC\,5530. The observations were obtained with the Australian Telescope Compact Array~(ATCA) more than 18\,yr after the explosion as part of the Rebrightening in Interacting Supernova Emission~(RISE) program, which monitors nearby core-collapse supernovae for late-time interaction with dense circumstellar material. 
SN\,2007it was detected on 2026~April~8~(08:00–12:00 UTC) at 5.5\,GHz with a flux density of $3.30 \pm 0.13\,$mJy and at 9.0\,GHz with $3.54 \pm 0.24\,$mJy. Its non-detection in publicly available $0.88$\,GHz ASKAP data from 2026~January~11 suggests either rapidly rising emission or significant internal absorption at lower frequencies. 
We assess the prospects for detection at other wavelengths and encourage coordinated follow-up observations across the radio, optical, X-ray, and $\gamma$-ray bands.

\end{abstract}


\section{Interacting and rebrightening Supernovae} 
 
Nearby core-collapse supernovae~(CC-SNe) are typically detected at radio frequencies within weeks to years after explosion, with the exact timescale depending on the SN subtype. SNe embedded in dense circumstellar material~(CSM) generally peak at later times~\citep[e.g., Type\,IIP and Type\,IIn SNe; ][]{2021ApJ...908...75B}. Infrared surveys have revealed unexpectedly bright emission from several SNe years after explosion~\citep{2013AJ....146....2F}. More recently, a growing number of SNe have been found to exhibit late-time radio rebrightening, with flux densities significantly exceeding expectations from standard light curves, declining $\propto t^{-1.5}$~\citep{2024MNRAS.534.3853R}.

Such rebrightening events are likely associated with interactions between the SN shock and dense circumstellar shells shaped by the progenitor’s mass-loss history, producing enhanced emission across multiple wavebands~\citep{2026A&A...706A..82B}. These environments may also provide favourable conditions for efficient acceleration particles~\citep{2025A&A...699A.160B}. However, most previous radio detections of late-time rebrightening were either serendipitous discoveries in archival data~\citep{2024MNRAS.534.3853R} or reported with substantial delays~\citep{2025PASA...42...50S}, precluding coordinated multiwavelength follow-up.

The Rebrightening in Interacting Supernova Emission~(RISE) initiative is designed to address this gap by systematically monitoring nearby CC-SNe from radio to X-ray energies and triggering $\gamma$-ray follow-up observations for the most promising events.

\section{ATCA and ASKAP Observations} 

As part of the RISE initiative, we observed the Type\,II supernova SN\,2007it in NGC\,5530 with the Australia Telescope Compact Array~(ATCA) using the BIGCAT backend on 2026~April~8~(08:00–12:00 UTC) under proposal C3827~(PI: R. Brose). The observations were conducted in the 6D array configuration. The ATCA primary calibrator PKS\,B1934--638 was observed at the end of the run to set the absolute flux density scale and correct for the frequency-dependent instrumental bandpass. The nearby secondary calibrator PKS\,1349--439 was observed for 2 minutes every 12 minutes to monitor and correct for variations in gain and phase. The data were reduced using standard tasks in the MIRIAD package~\citep{1994AAS..108..585S}, and images were produced using Briggs robust weighting of 0. 

\begin{figure}[!ht]
    \centering
         \includegraphics[width=\textwidth]{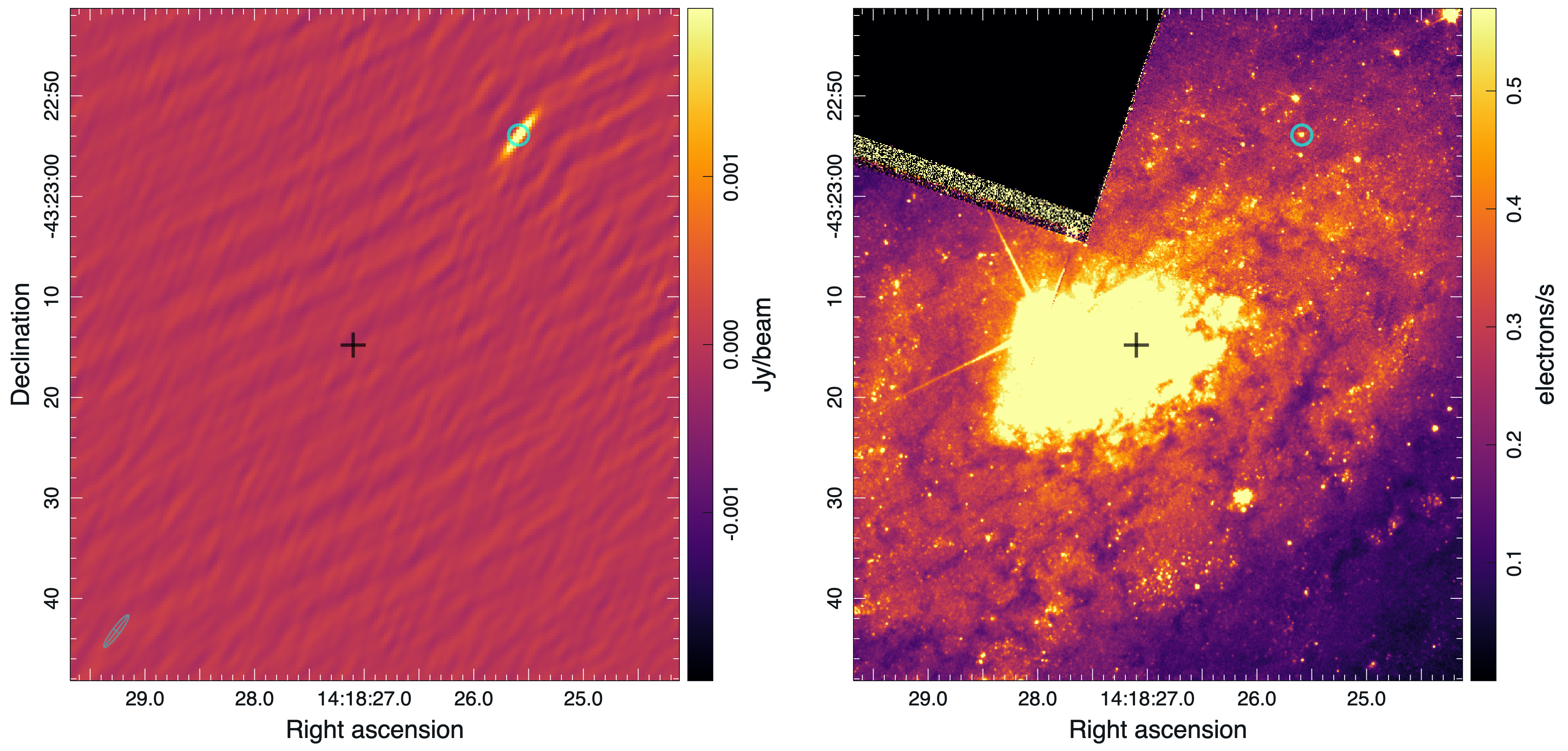}
    \caption{Left: ATCA 9\,GHz image showing the detection of SN\,2007it (cyan circle) from 2026\,April\,8. The centre of the host galaxy NGC\,5530 is marked with a black cross. The grey ellipse in the lower left corner represents a synthesised beam of $4.0''$$\times$\,$0.8''$. Right: Hubble Space Telescope image of NGC\,5530 from 2009\,May\,10, retrieved from the Hubble Legacy Archive.} 
    \label{fig:ATCA}
\end{figure}

SN\,2007it was detected~(Figure~\ref{fig:ATCA}) at 5.5\,GHz with a flux density of $3.30 \pm 0.13\,$mJy and at 9.0\,GHz with $3.54 \pm 0.24\,$mJy, corresponding to a spectral index of $\alpha=0.1\pm0.2$. 

We also performed forced photometry on publicly available ASKAP data at the SN position, deriving $3\sigma$ upper limits of $0.67\,$mJy~(2021~January~26) and $0.77\,$mJy~(2024~December~6), both at $1.3\,$GHz. These constraints are based on 15\,min integrations from RACS data~\citep{2023PASA...40...34D}. Multiple additional ASKAP observations were obtained between and after these dates, but none provided stronger constraints. The most recent epoch yields an upper limit of $3.1\,$mJy at $0.88\,$GHz from a 12\,min VAST observation~\citep{2021PASA...38...54M} on 2026~January~11.

\section{Discussion and Conclusion}

The radio emission from SN\,2007it is comparable to that of SN\,2001ig \citep{2025PASA...42...50S}. At 5.5\,GHz, the inferred radio luminosities are $4.1\times10^{26}\,$erg\,s$^{-1}$ Hz$^{-1}$ for SN\,2001ig and $5.9\times10^{26}\,$erg\,s$^{-1}$ Hz$^{-1}$ for SN\,2007it. SN\,2001ig exhibited the onset of radio rebrightening $\sim$12\,yr post-explosion, compared to $\sim$18\,yr for SN\,2007it, and remained bright for at least $\sim$11\,yr, with indications of variability. For SN\,2007it, the duration of elevated emission prior to 2026~April~8 is unconstrained. The non-detection in ASKAP data from 2026~January~11 may indicate either a rapid rise in radio luminosity or suppression of low-frequency emission due to internal absorption, the latter implying the presence of a substantial CSM. Further, careful analysis and comparison is needed in this regard though, as there is stronger emission from NGC\,5530 at lower frequencies as well as an increased beamsize compared to the reported ATCA observations.

Estimating the mass of the interacting CSM is non-trivial. For SN\,2001ig, shell masses of $M_{\rm shell}\sim1$–$4,M_\odot$ have been inferred, suggesting a comparable range for SN\,2007it. The only currently available models of particle acceleration in interacting SNe beyond simple scaling consider shell masses of $0.6M_\odot$ and $4.3M_\odot$~\citep{2026A&A...706A..82B}. Assuming a linear scaling between shell mass and radio luminosity for the peak-luminosities of the two models presented in Fig. 6 of \citet{2026A&A...706A..82B} yields
\begin{align}
 M_{\rm shell} &= (L_\text{\rm radio}-7.25\times10^{25}\text{erg/s}) \cdot 4.5\times10^{-27}\frac{M_\odot}{\text{erg/s}}+0.58M_\odot,
\end{align}
which implies $M_{\rm shell}\approx2.3M_\odot$ for SN\,2001ig and $\approx3M_\odot$ for SN\,2007it.

For such shell masses, detectable signatures at other wavelengths are expected. Shock–CSM interaction should produce thermal X-ray emission at levels up to $F_{0.2-10,{\rm keV}}\lesssim5.6\times10^{-11}\,$erg cm$^{-2}$ s$^{-1}$, optical emission in the $R$ band with $m\gtrsim23$ due to partial reprocessing of X-rays in the unshocked shell, and potentially detectable high- and very-high-energy $\gamma$-ray emission from enhanced $pp$ interactions, with fluxes of $F_{1-300,{\rm GeV}}\sim6\times10^{-12}\,$erg cm$^{-2}$ s$^{-1}$ and $F_{1-10,{\rm TeV}}\sim1\times10^{-12}\,$erg cm$^{-2}$ s$^{-1}$. Note, that those values are for the $4.3M_\odot$ shell from \citet{2026A&A...706A..82B} and the expected emission from SN2007it is weaker.


A growing number of SNe exhibit late-time radio rebrightening, although the origin of this phenomenon remains uncertain. Interaction of the SN shock with a dense, structured CSM provides a natural explanation and predicts multiwavelength counterparts. If a binary interaction produced a complex CSM configuration with enhanced magnetic fields, the expected signatures at other wavelengths may be significantly weaker. Coordinated and timely observations across the electromagnetic spectrum are therefore essential to constrain the origin of the rebrightening in SN\,2007it and similar events.

\begin{acknowledgments}
The ATCA is part of the Australia Telescope National Facility, which is funded by the Commonwealth of Australia for operation as a National Facility managed by CSIRO. We acknowledge the Gomeroi people as the traditional owners of the Observatory site. 
This scientific work uses data obtained from Inyarrimanha Ilgari Bundara / the CSIRO’s Murchison Radio-astronomy Observatory. We acknowledge the Wajarri Yamaji People as the Traditional Owners and native title holders of the Observatory site. CSIRO’s ASKAP radio telescope is part of the Australia Telescope National Facility (\url{https://ror.org/05qajvd42}). Operation of ASKAP is funded by the Australian Government with support from the National Collaborative Research Infrastructure Strategy. ASKAP uses the resources of the Pawsey Supercomputing Research Centre. Establishment of ASKAP, Inyarrimanha Ilgari Bundara, the CSIRO Murchison Radio-astronomy Observatory and the Pawsey Supercomputing Research Centre are initiatives of the Australian Government, with support from the Government of Western Australia and the Science and Industry Endowment Fund. Archived data can be obtained through the CSIRO ASKAP Science Data Archive, CASDA (\url{http://data.csiro.au/}). 
\end{acknowledgments}

\begin{contribution}
S. Lazarevi\'c, K. Rose, S. Ryder, Z. J. Smeaton, R. Z. E. Alsaberi and C. Burger-Scheidlin conducted the RISE ATCA pilot survey observations. R. Z. E. Alsaberi, Z. J. Smeaton, S. Lazarevi\'c and M. Filipovi\'c performed the ATCA data reduction and analysis. K. Rose and S. Lazarevi\'c extracted limits from the ASKAP observations. R. Brose provided the multi-wavelength modelling. All authors contributed equally to the writing of the manuscript. 
\end{contribution}

\bibliography{Note}{}
\bibliographystyle{aasjournalv7}

\end{document}